\documentclass[12pt,preprint]{aastex}

\slugcomment{LA-UR-06-1067}
\shorttitle{Pseudo-continuum opacity in WD atmospheres}
\shortauthors{Kowalski}

\def\wig#1{\mathrel{\hbox{\hbox to 0pt{%
          \lower.5ex\hbox{$\sim$}\hss}\raise.4ex\hbox{$#1$}}}}

\begin{document}

\title{The Pseudo-continuum Bound-free Opacity of Hydrogen and its Importance 
in Cool White Dwarf Atmospheres}
\author{Piotr M. Kowalski}

\affil{Department of Physics and Astronomy, Vanderbilt University, Nashville, TN 
37235-1807}
\affil{Los Alamos National Laboratory, MS F699, Los Alamos, NM 87545}
\email{kowalski@lanl.gov}

\begin{abstract}

We investigate the importance of the pseudo-continuum bound-free opacity from hydrogen atoms in the atmospheres of cool white dwarfs. 
This source of absorption, when calculated by the occupation probability formalism applied in the modeling 
of white dwarf atmospheres with $T_{\rm eff}\rm <17000 \, K$, dominates all other sources of opacity at optical wavelengths. 
This is unrealistic and not observed. On the other hand, a significant flux suppression in the blue part of the spectra 
of cool white dwarfs has been reported, and mainly interpreted as a result of the pseudo-continuum absorption from atomic hydrogen. 
We investigate this problem by proposing 
a new, more realistic approach to calculating this source of opacity. We show that this absorption is orders of magnitude 
smaller than that predicted by current methods. Therefore, we rule out the pseudo-continuum opacity as a source of the flux deficiency 
observed in the spectra of cool white dwarfs.

\end{abstract}

\keywords{atomic processes -- dense matter -- stars: atmospheres -- stars: white dwarfs}

\section{Introduction}

Cool white dwarfs are among the oldest objects in the Universe. The detection of 
large populations of these old stars in our Galaxy gives us a unique opportunity to use them as 
chronometers. To achieve this goal, we need reliable atmosphere models for cool white dwarfs. 
Such models would allow for a better determination of the atmospheric parameters of observed white dwarf stars, 
tracking the evolution of their atmospheric composition, and to obtain better boundary conditions for modeling internal 
structures and the cooling of these stars \citep{FN00}. 
One problem in modeling cool white dwarf atmospheres
is the implementation of the D\"appen-Anderson-Mihalas (hereafter DAM) model \citep{DAM} 
in calculating the pseudo-continuum bound-free opacity from the ground state of atomic hydrogen. 
When this source of opacity is calculated following the DAM prescription with the occupation probabilities of Hummer \& Mihalas (1988, hereafter HM)
for white dwarfs with $T_{\rm eff}\rm<17000 \, K$, it dominates all other absorption processes at optical wavelengths. 
Such strong absorption is completely unrealistic and is not observed \citep{Bergeron01,Bergeron97}. 
Nonetheless, this process may still be significant in white dwarf atmospheres. In fact, \citet{Bergeron97} 
suggested that this absorption mechanism may be responsible for the observed flux suppression in the blue part of the spectra of cool 
white dwarfs. Following that idea, \citet{Bergeron01} introduced an arbitrary damping 
function to the DAM pseudo-continuum opacity to obtain good fits of the models to the photometry of selected stars. 
However, the damping function must be adjusted for each star, indicating how unsuitable the DAM model is. As a result, 
the pseudo-continuum bound-free opacity of hydrogen atoms is ignored in the modeling of cool white dwarf atmospheres.
In this paper we readdress this issue by computing this absorption mechanism with a much more realistic physical model
to assess its importance in cool white dwarf atmospheres.
  
The pseudo-continuum bound-free opacity results 
from the perturbation of hydrogen atoms in their ground state by their interaction with other particles in a dense medium \citep{DAM}. 
Such perturbations result in a lowering of the ionization barrier of some of the hydrogen atoms 
and the possibility of a bound-free transition caused by photon with energies that are smaller than the ionization potential 
of the isolated hydrogen atom. DAM consider this process in the framework of the occupation probability formalism proposed by HM. 
They constructed a simple model for the optical properties of an interacting hydrogen medium. In the weakly ionized 
atmospheres of cool white dwarfs, the hydrogen atoms in higher excited states are perturbed mostly by neutral particles 
\citep{BWF91} and eventually destroyed (pressure ionized) by excluded volume interactions. 
The effect is stronger for excited states because the size of a hydrogen atom
is a monotonically increasing function of the principal quantum number. 
This model for the pseudo-continuum opacity is not very precise, as it treats the inter-particle interactions 
and the associated lowering of the ionization potential only in terms of the average sizes of the particles. 
We will show that a quantum mechanical description of the interactions between hydrogen atoms and 
perturbers, absent in the simple DAM model, is necessary to obtain a physically realistic value of 
the pseudo-continuum opacity of hydrogen atoms in the atmospheres of white dwarf stars.

\begin{figure*}[t]
\centering
\epsscale{0.65}
\plotone{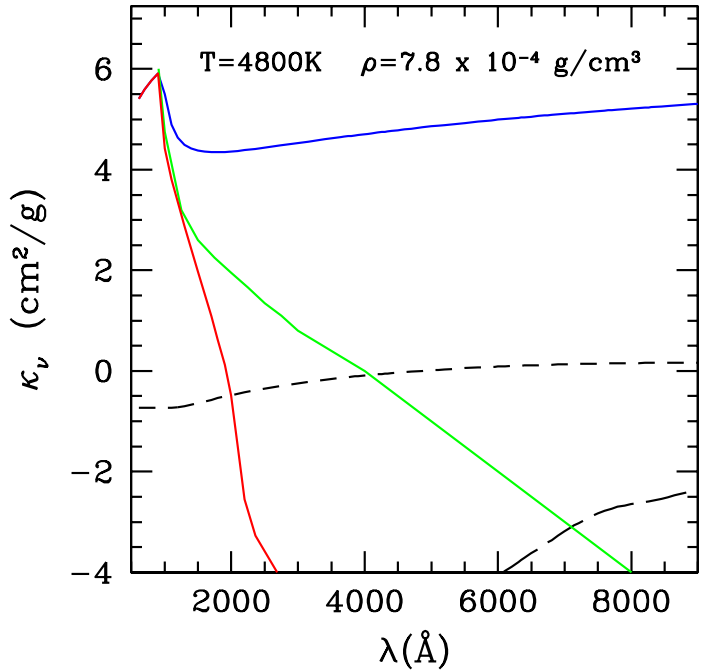} 
\figcaption[f1.ps]{\scriptsize The most relevant sources of opacity at the photosphere of a pure hydrogen model atmosphere with
$T_{\rm eff}\rm =4600K$ and $\rm log \it \, g\rm=7.75 \, (cgs)$ - the atmospheric parameters of the white dwarf star LP 380-5 presented on figure 4 
of \citet{Bergeron01}. The lines represent the pseudo-continuum opacity presented here (red solid), the pseudo-continuum opacity of DAM (blue solid), 
the $\rm H^{-}$ bound-free opacity (dashed), the $\rm H_2$ CIA opacity (long dashed) and the pseudo-continuum opacity derived by \citet{Bergeron01} 
to fit the spectrum of this star (green solid).
       \label{F1}}
%\figcaption{Fig 1.}
\end{figure*}

To accurately account for the effect of lowering the hydrogen ionization barrier in a dense medium, we consider 
short-range, binary collisions. The ionization energy of the perturbed hydrogen atom is given by the ionization energy 
of the H-perturber pair. This energy is given by the difference between the exact interaction energy curves 
calculated quantum mechanically of the neutral and 
singly ionized dimers (interacting pairs). The probability of having 
a pair colliding with a given collision distance can be determined from the interaction potential of the neutral dimer. 
The resulting probabilities of lowered hydrogen ionization potential for the physical conditions of white dwarfs atmospheres
are orders of magnitude smaller than those obtained from the DAM model with the HM occupation probabilities. 
%Our model indicates that the pseudo-continuum opacity from atomic hydrogen
%is unimportant in the modeling of cool white dwarf atmospheres.

In the next section we discuss the DAM approach to the pseudo-continuum opacity. 
Our new calculations of this absorption mechanism and its application to the physical conditions found 
in the atmospheres of cool white dwarfs are presented in sections 3 and 4, respectively.
 
\section{The pseudo-continuum opacity model of D\"appen-Anderson-Mihalas}
The DAM model is based on the occupation probability formalism introduced by HM
for calculating thermodynamical properties of a non-ideal gas. HM introduced the non-ideal effects on the equation of state by modifying 
the internal partition functions of bound species. The modified partition function of the hydrogen atom is given by
\begin{equation} Z_{\rm H}=\sum_{i}g_{i}\omega_{i}e^{-\epsilon_{i}/k_{B}T}=\sum_{i}\omega_{i}\gamma_{i},\end{equation}
where $k_{B}$ is the Boltzmann constant, $i$ indicates the atomic level, $g_i$ is the statistical weight, $\epsilon_{i}$ is the energy 
of hydrogen atom counted from the energy of the ground state ($i\rm =0$), $T$ is the temperature, and $\omega_i$
is the occupation probability. In the HM model, $\omega_{i}$ represents the strength of non-ideal gas perturbations on an atomic level \it i \rm. 
Its value for the physical conditions found inside weakly ionized cool white dwarf atmospheres is given by the excluded volume interactions model \citep{BWF91}
and decreases for higher principal quantum number (HM, Equation 3.4). 
The number of hydrogen atoms in the excited state $i$ is expressed as
\begin{equation} \frac{n_{i}}{n_{tot}}=\frac{\omega_{i}\gamma_{i}}{Z_{\rm H}}, \label{3} \end{equation}
where $n_{tot}$ is the total number density of atomic hydrogen.

DAM extended this model for the calculation of the optical properties of hydrogen atoms in the partially ionized plasma. 
They interpreted the factor $1-\omega_{i}$ as the fraction of hydrogen atoms with atomic level $i$ being ``dissolved," i.e. sufficiently perturbed to describe 
an unbound electron and ion. They assigned this value to the fraction of free-electrons states available for a bound-free transition to level $i$.
DAM used this interpretation to derive that for the absorption of a photon associated with the transition from level $i$ to level $j$, 
the probability that level $j$ is bound, and the transition is ``bound-bound," is $\omega_j/\omega_i<1$. Therefore, the probability of having
a ``bound-free" transition is
\begin{equation} P_{bf}=1-\frac{\omega_j}{\omega_i}.\end{equation}
To describe the pseudo-continuum absorption to the continuum energy levels localized between two discrete but ``dissolved" levels of the hydrogen atom, 
DAM introduced the pseudo-level $n^{*}$ with occupation probability $\omega_{n^{*}}$. The latter is calculated by the interpolation 
between the $\omega_{i}$'s. Therefore, 
the probability that, during the $i$ to $n^{*}$ transition, the absorbing electron goes into the continuum is given by (Eq. 32 of DAM)
\begin{equation} P_{bf}=1-\frac{\omega_{n^{*}}}{\omega_i}. \label{9a} \end{equation}

\begin{figure*}[t]
\centering
\epsscale{0.8}
\plotone{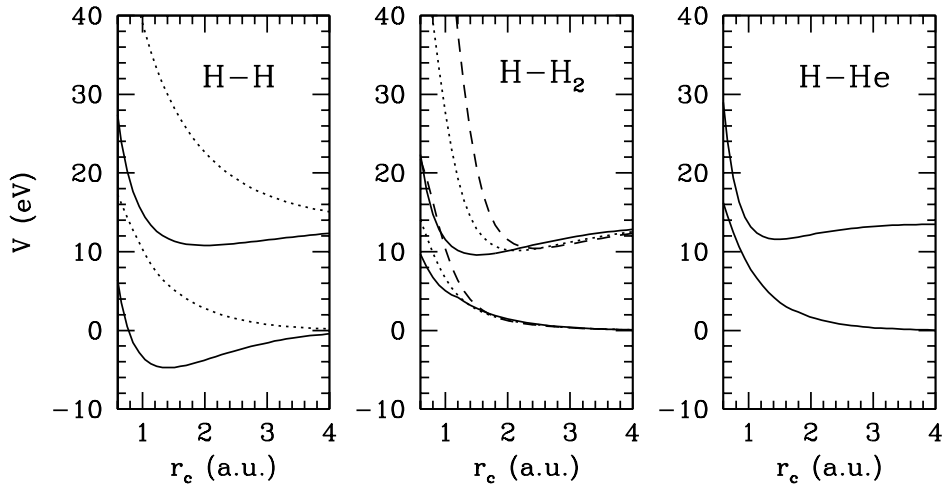} 
\figcaption[f2.ps]{\scriptsize
The interaction energy curves as a function of the inter-particle collision distance for neutral dimers (lower curves) and the corresponding 
single-ionized dimers (upper curves). The different curves for the $\rm H-H$ interaction are for the \it bound \rm (solid line) 
and \it anti-bound \rm (dotted line) interaction potentials for $\rm H-H$ and $\rm H_2^{+}$. The different energy curves for the $\rm H-H_2$ 
interaction represent the potentials for different orientation of the molecule, with the angle defined between the line connecting 
the perturber to the center of $\rm H_2$ and the molecular axis. Interaction for the following angles are shown: $\rm 90^{o}$ (solid), 
$\rm 45^{o}$ (dotted), and $\rm 0^{o}$ (dashed). 
       \label{F2}}
%\caption{Fig 2.}
\end{figure*}

The dominant perturbers in the atmospheres of cool white dwarfs are neutral particles. Therefore in calculating the DAM pseudo-continuum opacity
we used the hard sphere model for $\omega_i$ of HM (Section IIIa). In Figure 1 we present the main opacity sources at the photosphere of white dwarf star
LP 380-5 considered for the potential presence of pseudo-continuum opacity by \citet{Bergeron01}. The DAM pseudo-continuum opacity is much too high
and dominates all other opacity sources by orders of magnitude. This is unrealistic and simply not observed.
Therefore, the DAM model highly overestimates the probabilities of pseudo-continuum bound-free absorption (Eq. \ref{9a}) in the atmospheres 
of cool white dwarf stars. In view of the difficulty with this application of the DAM model of the pseudo-continuum opacity, we propose 
a more realistic approach to calculate this source of absorption.

\section{A new model for the pseudo-continuum opacity of H}
The bound-free absorption process results in the ionization of a hydrogen atom. The lowering of the hydrogen ionization potential 
arises from the interaction of the hydrogen atom with neighboring particles \citep{DAM}. 
However, we are interested in the pseudo-continuum opacity far from the Lyman edge, i.e. $\rm \lambda\wig>1500 \, \AA$.
A significant lowering of the hydrogen ionization barrier, by more than $\sim 5 \, \rm eV$, is required for bound-free absorption 
at these wavelengths. Such a situation occurs in the case of rare, close collisions for which the inter-particle collision distances, $r_{c}$, 
are small enough that the probability ($P_c$) of finding a colliding pair with an inter-particle separation smaller than $r_{c}$ is much smaller than unity.
In this case, multi-particle collisions are insignificant, as the probability of having a close $N$-particle collision is roughly $\sim P_c^{N-1}$. 
Therefore, it is sufficient to consider the interaction between a $\rm H$ atom and its closest neighbor only. \citet{Allard} 
used this approximation to successfully explain the complex shape of the Lyman $\rm \alpha$ line wings detected in
the spectra of white dwarfs with $T_{\rm eff}\rm \sim 12000 \, K$.

For a given colliding pair the change in the ionization energy results from the formation of a temporary dimer, whose ionization energy differs from that
of the isolated hydrogen atom,
$I_0\rm =13.598 \, eV$. This modified ionization barrier $I_p$ is given by the ionization energy
of a dimer calculated at a fixed inter-particle separation $r_{c}$:
\begin{equation} I_p(r_c)=V_{\rm H^+-pert}(r_c)-V_{\rm H-pert}(r_c),\end{equation}
where $V_{\rm H-pert}(r_c)$ and $V_{\rm H^+-pert}(r_c)$ are the energies of the neutral and singly ionized dimers, respectively.
This picture is in the spirit 
of the Franck-Condon principle, which states that as a result of large differences between the mass of the absorbing electron and the nuclear mass, 
the nuclei remain fixed during the absorption/emission of a photon \citep{Davydov}. The differential probability of finding such a dimer with an inter-particle 
separation between $r_c$ and $r_c+dr_c$ in the low density medium ($\rho \wig< 0.1 \rm \, g/cm^3$) is given by \citep{MS}. 

\begin{figure*}[t]
\centering
\epsscale{0.8}
\plotone{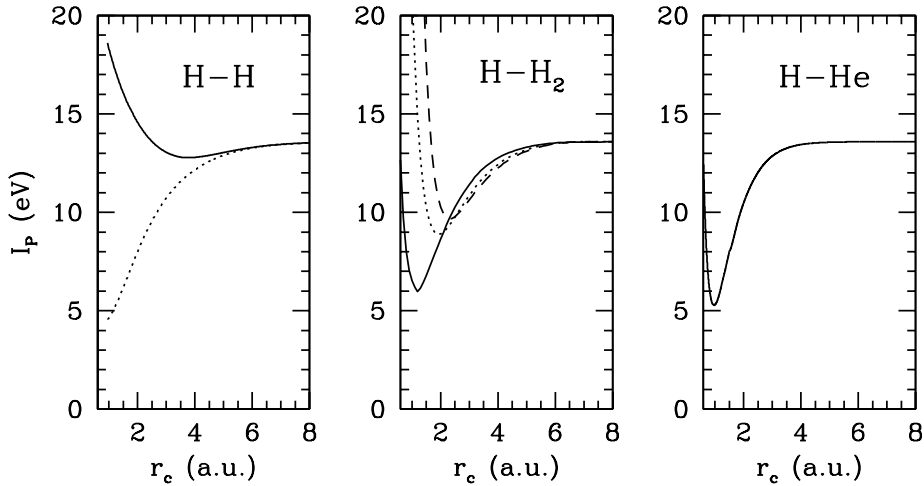} 
%\caption{Fig 3.}
\figcaption[f3.ps]{\scriptsize
The ionization energy of dimers as a function of the inter-particle distance. See Fig. 2 for legend.
       \label{F3}}
\end{figure*}

\begin{equation} dP(r_c)=n_{pert}\left(\int_{\theta,\phi}{e^{-V_{\rm H-pert}(r_c,\theta,\phi)/k_BT} \sin \theta d \theta d \phi}\right)r_c^2dr_c , \label{11}\end{equation}
where $n_{pert}$ is a number density of perturbers, $V_{\rm H-pert}$ is the interaction energy between a hydrogen atom and the perturber localized at the position $(r_c,\theta,\phi)$
in relation to the hydrogen atom, which is assumed to be at the origin of the coordinate system. 
The dominant species, and therefore the main perturbers, in the atmospheres 
of cool white dwarfs are $\rm H$, $\rm H_2$, and $\rm He$. The sources for the potentials for the interaction of hydrogen atom with perturbers are chosen 
to be: $\rm H-H$ [\citet{KW}], $\rm H-H_2$ [\citet{BOO}], and $\rm H-He$ [\citet{SH}]. The corresponding potential curves for the singly ionized dimers are
chosen to be:
$\rm H_2^{+}$ [\citet{BR}], $\rm H_3^{+}$ [\citet{PROS}], and $\rm HeH^{+}$ [\citet{GR}]. These potentials are plotted on Fig. 2. As a result of the 
\it gerade/ungerade \rm symmetry in the $\rm H-H$ interaction, we have to consider both the \it bound \rm and the \it anti-bound \rm $\rm H-H$ potential energy curves. 
%This consideration is because the type of interaction for a given pair depends on the relative orientation of the electronic spins. 
For the singly ionized dimers we choose the ground state potential energy curves because
the upper energy curves (like \it anti-bound \rm state for $\rm H_2^+$, Fig. 2.) would result in a much smaller decrease in the ionization energy for a given $r_c$. 
This decrease occurs with much smaller probability than the same change in the ionization potential that results from the ionic ground states at a much larger $r_{c}$, and therefore can 
be neglected. The resulting ionization energy $I_p(r_c)$ for a given dimer as a function of the collision distance are plotted on Fig. 3. The ionization energy
can decrease to $\rm \sim 5 \, eV$, which occurs at very small $r_{c}$. Due to the strong repulsion in the neutral dimer at short separation (Fig. 2), 
such a large lowering of the ionization energy occurs with very small
probability, as for small $r_c$ the Boltzmann factor $e^{-V_{\rm H-pert}/k_BT}<<1$ in Eq. (\ref{11}). Moreover, in the case of the $\rm H-H$ interaction, 
only the \it anti-bound \rm potential leads to a significant decrease of the ionization barrier (Fig. 2).  

\begin{figure*}[t]
\centering
\epsscale{0.8}
\plotone{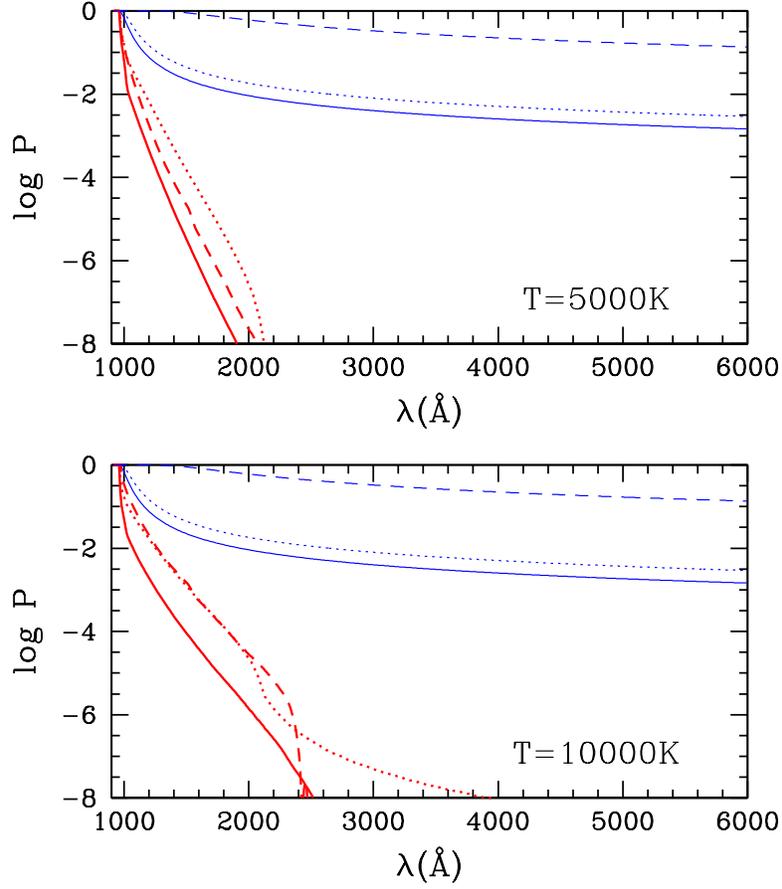} 
\figcaption[f4.ps]{\scriptsize The probability of lowering the bound-free continuum from our model (thick red lines), and
the DAM approach (thin blue lines), as a function of wavelength for selected temperatures. Calculations for the compositions are shown: 
(1) $n_{\rm H}\rm =10^{21} \ cm^{-3}$ (solid line), (2) $n_{\rm H}=n_{\rm H_2}\rm =10^{21} \ cm^{-3}$ (dotted line), and 
(3) $n_{\rm He}\rm =10^{23} \ cm^{-3}$ (dashed line).
       \label{F4}}
%\caption{Fig 4.}
\end{figure*}

The probability of a hydrogen atom having an ionization energy $I_p$ sufficiently smaller for photo-ionization to be caused by a photon 
of frequency $\nu$ is
\begin{equation} P(\nu)=\int_{I_p(r_c)<h\nu} dP(r_c),\end{equation}
where the integration is performed over the range of separations $r_c$ such that $I_p(r_c)<h\nu$. 
The resulting probabilities, as a function of photon wavelength $\rm \lambda$, of having a bound-free transition 
for $\rm H$ for two temperatures and three different compositions are shown in Fig. 4. The corresponding probabilities from the DAM model
(also shown in Fig. 4) are a few orders of magnitude larger. We find that for a given density, collisions with $\rm H_2$ are most effective at
lowering the ionization energy of $\rm H$. This is because for a given value of $I_p(r_c)$, the interaction potential $\rm V_{H-H_2}(r_c)$ computed 
for the orientation angle $\sim 90^{o}$ (see caption of Fig 2. for definition) is much less repulsive than $\rm V_{H-H}(r_c)$ and $\rm V_{H-He}(r_c)$.  

\begin{figure*}[t]
\centering
\epsscale{0.8}
\plotone{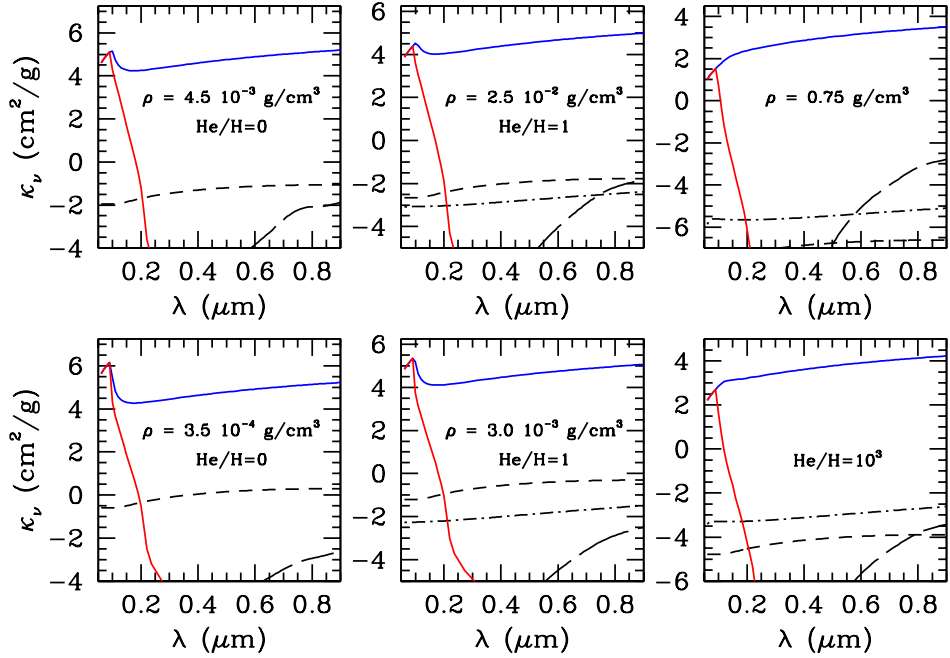} 
%\caption{Fig 4.}
\figcaption[f5.ps]{\scriptsize The most relevant sources of opacity for representative physical conditions at the photosphere of cool white dwarf 
stars. Upper panels: $T\rm=4000 \, K$, lower panels: $T\rm=5000 \, K$.
The lines represent the pseudo-continuum opacity calculated here (red solid), the pseudo-continuum opacity of DAM (blue solid), the $\rm H^{-}$ bound-free opacity (dashed),
the $\rm H_2$ CIA opacity (long dashed), and the $\rm He^{-}$ free-free opacity (dash-dotted).
       \label{F5}}
\end{figure*}

The pseudo-continuum opacity is then obtained with
\begin{equation} \kappa_{bf}(\nu)=\frac{n_H}{\rho} P(\nu)\sigma^0_{bf}(\nu) \ \ \ (h\nu< \rm 13.598 \, eV)\label{333} \end{equation} 
where $\sigma^0_{bf}(\nu)$ is the bound-free cross-section of the isolated hydrogen atom extrapolated beyond the Lyman edge (DAM), $n_{\rm H}$ is 
the number density of hydrogen atoms, and $\rho$ is the mass density.
A proper calculation would consider the bound-free cross-section of the dimer as a function of inter-particle separation.
To our knowledge this information is not available. However, as the photo-ionization cross-sections for $\rm H$ \citep{PL} 
and $\rm H_2$ \citep{FORD} differ by less than a factor of $3$ and
the cross-section is the effective size of the absorbing system as seen by the photon, we estimate that the extrapolation of the bound-free cross-section 
of the isolated hydrogen atom beyond Lyman edge introduces an uncertainty no larger than an order of magnitude on $\kappa_{bf}(\nu)$ calculated by Eq. \ref{333}. 
In helium-rich atmospheres, where the density can be as high as $\rm 2-3 \, g/cm^3$, equation (\ref{11}) should be corrected when $\rho \rm \wig>0.1\,g/cm^3$, 
by a factor $e^{\rm w_{\rm H-pert}}$ \citep{MS},
where $\rm w_{\rm H-pert}$ is a thermal potential that arises from the correlations in the dense fluid. 
Solving the Ornstein-Zernike equation in the Percus-Yevick approximation \citep{MS} for fluid $\rm He$, we have verified that in the helium-rich
atmospheres of cool white dwarfs, 
$e^{\rm w_{\rm H-pert}}<100$. Therefore, due to correlations in the most extreme case, Eq. \ref{333} leads to an underestimate of $\kappa_{bf}$ of
no more than two orders of magnitude. We will see below that this does not affect our conclusions.

\section{Importance of the pseudo-continuum opacity in cool white dwarf atmospheres}
Our goal is to investigate the suggestion of \citet{Bergeron97} and \citet{Bergeron01} that the pseudo-continuum bound-free absorption
by atomic hydrogen beyond the Lyman edge may be the missing source of opacity in the atmosphere models of cool white dwarfs $(T_{\rm eff}\rm \wig< 6000 \, K)$. 
For this purpose, we have computed opacities for temperatures, densities and $\rm He/H$ composition that are representative of the photospheres of these stars. 
%We assumed 
%the surface gravity of $\rm log \, \it g \rm = 8$, and various $\rm He/H$ ratios. 
%The new pseudo-continuum bound-free opacity is presented on Fig. 5. 
The pseudo-continuum opacity calculated with our model dominates all other sources 
of opacity at wavelengths shorter than $\rm \sim 2000\,\AA$ (Fig. 5). 
For these cool stars the flux at these wavelengths is extremely small and completely negligible (see Fig. 5 of \citet{Bergeron01}).
We also note that the uncertainty in the photo-ionization cross-section and our neglect of the correlations in dense helium-rich atmospheres 
do not alter the pseudo-continuum opacity enough to make it important at $\rm \lambda>2000\,\AA$. 

Fig. 1. reproduces the opacity plot of \citet{Bergeron01} (Fig. 4 in that paper). This figure represents the main sources of opacity    
for the physical conditions found at the photosphere of the white dwarf star LP 380-5. Our pseudo-continuum opacity is several orders of magnitude 
smaller than that necessary to fit the blue spectrum ($\rm 3000-4000\,\AA$) of this star \citep{Bergeron01}, and is completely insignificant at $\rm \lambda>2000 \, \AA$.
%Therefore the suspected absorption mechanism in the blue spectral region of cool white dwarf stars can not be attributed to this source of opacity. 

\section{Conclusions}
The possible presence of an unknown absorption mechanism in the atmospheres of cool white dwarfs has been reported by \citet{Bergeron01} and \citet{Bergeron97}. 
This opacity source has been attributed to the pseudo-continuum bound-free opacity from hydrogen atoms in their ground state that results from a lowering 
of the ionization potential because of inter-particle interactions in the gas. Opacity models based 
on the occupation probability formalism highly overestimate this source of absorption in the atmospheres of cool white dwarfs. For this reason, 
it is usually omitted in models. 
In this paper, we presented a realistic model for this absorption mechanism based on binary collisions, exact pair interaction potentials, 
and the ionization energy of the colliding pair. We find that the pseudo-continuum bound-free opacity decreases very rapidly with wavelength beyond 
the Lyman edge and becomes completely negligible beyond $\lambda \sim 2000 \, \rm \AA$ over the entire range of temperature, density and $\rm H/He$ composition
relevant to cool white dwarf atmospheres. Therefore, another absorption mechanism must be invoked to explain the flux excess of the models in the blue part 
of the spectrum of these stars.

I thank D. Saumon for useful discussions and comments on this manuscript.
This research was supported by the United States Department of Energy under contract W-7405-ENG-36.

\clearpage


\small
   
\begin{thebibliography}{}
\bibitem[Allard et al., 2004]{Allard} Allard, N. F., Kielkopf, J. F., \& Loeillet, B. 2004, \aap, 424, 347
\bibitem[Bates \& Reid, 1968]{BR} Bates, D. R., \& Reid, H. G. 1968, Adv. Atom. and Molec. Phys. 4, 13
\bibitem[Bergeron, 2001]{Bergeron01} Bergeron, P. 2001, \apj, 558, 369
\bibitem[Bergeron et al., 1997]{Bergeron97} Bergeron, P., Ruiz, M. T., \& Leggett, S. K. 1997, \apj, 108, 339
\bibitem[Bergeron et al., 1991]{BWF91} Bergeron, P., Wesemael, F., \& Fontaine, G. 1991, \apj, 367, 253
\bibitem[Broothroyd et al., 1991]{BOO} Boothroyd, A. I., Keogh, W. J., Martin, P. G., Peterson, M. R. 1991, J. Chem. Phys., 95, 4343
\bibitem[D\"appen et al., 1987]{DAM} D\"appen, W., Anderson, L. \& Mihalas, D. 1987, \apj, 319, 195 (DAM)
\bibitem[Davydov, 1965]{Davydov} Davydov, A. S. 1965, Quantum Mechanics (Oxford: Addison-Wesley), Chap. 12, Sec. 123
\bibitem[Fontaine et al., 2000]{FN00} Fontaine, G., Brassard, P. \& Bergeron, P. 2000, PASP, 113, 409 
\bibitem[Ford et al., 1974]{FORD} Ford, A. L., Kate, K. D. \& Dalgarno, A. 1975, \apj, 195, 819
\bibitem[Green et al., 1974]{GR} Green, T. A. et al. 1974, J. Chem. Phys., 61, 5186
\bibitem[Hummer \& Mihalas, 1988]{HM} Hummer, D. G., \& Mihalas, D. 1988, \apj, 331, 794 (HM)
\bibitem[Kolos \& Wolniewicz, 1965]{KW} Kolos, W., \& Wolniewicz, L. 1965, J. Chem. Phys. 43, 2429
\bibitem[Martynov, 1992]{MS} Martynov, G. A. 1992, Fundamental theory of liquids (Bristol: Adam Hilger), Chap. 5
\bibitem[Palenius et al., 1975]{PL} Palenius, H. P., Kohl, J. L., \& Parkinson, W. H. 1975 Phys. Rev. A. 13, 1805
\bibitem[Prosmiti et al., 1997]{PROS} Prosmiti, R.  Polyansky, O. L. and Tennyson, J. 1997, Chem. Phys. Lett. 273, 107
\bibitem[Shalabi et al., 1998]{SH} Shalabi, A. S., Eid, Kh. M., Kamel, M. A., \& El-Barbary, A. A. 1998, Physics Letters A, 239, 87


\end{thebibliography}
\end{document}